\documentclass[11pt,twoside]{article}

\usepackage{asp2006}
\usepackage{epsf}
\usepackage{psfig}
\usepackage{lscape}

\begin{document}
\title{Iron Line Profiles from Relativistic Thick Accretion Disk} 
\author{Sheng-Miao Wu and Ting-Gui Wang}   
\affil{Center for Astrophysics, University of Science and
Technology of China, Hefei, 230026, China}    

\begin{abstract}
We present a new code for calculating the Fe K$\alpha$ line profiles
from relativistic accretion disks with finite thickness around a
rotating black hole. The thin Keplerian accretion disk must become
thicker and sub-Keplerian with increasing accretion rates. We here
embark on, for the first time, a fully relativistic computation
which is aimed at gaining an insight into the effects of geometrical
thickness and the sub-Keplerian orbital velocity on the line
profiles. This code is also well-suited to produce accretion disk
images.
\end{abstract}


\section{Introduction}   
The fluorescent K$\alpha$ iron emission line has been observed in
several active galactic nuclei(AGN). This line is thought to be
produced by iron fluorescence of a relatively cold X-ray illuminated
material in the innermost parts of the accretion disk. Thus the line
profile encodes the nature of the structure, geometry, and dynamics of
the accretion flow, as well as the geometry of the space.

    Calculations of the line profiles have been performed by several
authors. Two basic approaches have been described: 1) using a kind of
transfer function to map a disk into sky plane \citep{cun75,lao91,spe95};
2) a ray tracing approach \citep{cad98,bec04}. But all of these techniques
are restricted to the geometrically thin, Keplerian disk located in the 
equatorial plane. To our knowledge, the disk must become geometrically thick
and be sub-Keplerain with increasing accretion rates
\citep{wan99,sha05}. Motivated by the above considerations a
relativistic thick accretion disk model is presented.
\section{Assumptions and Results} 
For simplicity, we adopt a conical surface for the disk geometry. The 
thickness of the disk can be described by the declination angle $\delta$($0 \le
\delta \le \pi/4$). When $\delta = 0$, the disk reduced to a thin disk. For
sub-Keplerian velocity, we adopt the modification of $\Omega$ firstly
introduced by \citet{rus00}
\begin{equation}
\Omega=\left(\frac{\vartheta}{\pi/2}\right)^{1/n}\Omega_K+\left[1-\left(
\frac{\vartheta}{\pi/2}\right)^{1/n}\right]\omega. \label{omega}
\end{equation}
It is easy to verify that $\Omega\le\Omega_K,\,\Omega_K$ is the
Keplerian angular velocity. The parameters of this model include:
the radii of the emitting disk zone $r_{\rm in}, r_{\rm out}$; the
spin of the black hole $a$; the inclination angle of the
disk($\vartheta_{\rm o}$) and the angle($\vartheta_{\rm e}$) between
the rotation axis of the system and the radial direction of the disk
surface which satisfies $\delta+\vartheta_{\rm e}=\pi/2$; the radial
emissivity index $p$; the angular velocity index $n$, respectively.

\begin{figure}[!ht]
\plotone{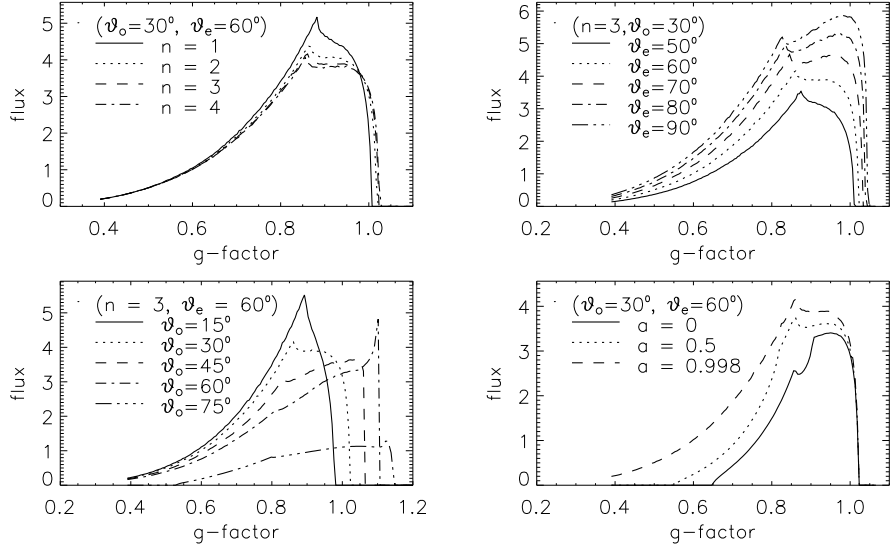} \vspace{-1.5cm} \caption{Examples of
line profiles showing the impact of different model parameters to the profile. 
The emission line region is from $r_{\rm ms}$
to $r_{\rm out}= 20 r_{\rm g}$, where $r_{\rm g}$ is the
gravitational radius. Top left: line profiles as a function of the
angular velocity for $n=1,2,3,4$. Top right: line profiles
as a function of the disk thickness for $\vartheta_{\rm e}
= 50^{\circ},60^{\circ},70^{\circ},80^{\circ},90^{\circ}$. Bottom
left: Comparison of the relativistic line profiles generated by our
model with different observer inclinations $\vartheta_{\rm o} =
15^{\circ},30^{\circ},45^{\circ},60^{\circ},75^{\circ}$. Bottom
right: Comparison of the relativistic line profiles generated by our
model with different spins $a=0,\, 0.5,\, 0.998$ and $n=3.$ The flux
in all cases is given using the same arbitrary units, and all our
results are unsmoothed.
 \label{lineprof}}
\end{figure}

    Following the idea presented by \citet{spe95}, we extend their model 
to the finite thick disk, and adopt elliptic integrals to improve the 
performance of the code. Here, we show, as an example, a number of line 
profiles obtained by our code(see Fig.\ref{lineprof}), the emissivity law 
in all cases is taken the forms $\epsilon(r_{\rm e}) \propto r_{\rm e}^{-3}$ 
and $f(\mu_{\rm e}) \propto (1+2.06 \mu_{\rm e})$.



\end{document}